\documentstyle[multicol,aps,prb,psfig]{revtex}
\begin{document}

\title {Dislocations in the ground state of the solid-on-solid model on a 
disordered substrate}

\author{Frank O. Pfeiffer$^{1}$ and Heiko Rieger$^{1,2}$}

\address{
$^{1}$ Institut f\"ur Theoretische Physik, Universit\"at zu K\"oln, 
       50937 K\"oln, Germany\\
$^{2}$ Institut f\"ur Theoretische Physik , Universit\"at des Saarlandes,
       66041 Saarbr\"ucken, Germany
}
\date{ \today}
\maketitle

\begin{abstract}
  We investigate the effects of topological defects (dislocations) to the
  ground state of the solid-on-solid (SOS) model on a simple cubic disordered
  substrate utilizing the min-cost-flow algorithm from combinatorial
  optimization.  The dislocations are found to destabilize and destroy the
  elastic phase, particularly when the defects are placed only in partially
  optimized positions.  For multi defect pairs their density decreases
  exponentially with the vortex core energy.  Their mean distance has a
  maximum depending on the vortex core energy and system size, which gives a
  fractal dimension of $1.27 \pm 0.02$.  The maximal mean distances correspond
  to special vortex core energies for which the scaling behavior of the
  density of dislocations change from a pure exponential decay to a stretched
  one.  Furthermore, an extra introduced vortex pair is screened due to the
  disorder-induced defects and its energy is linear in the vortex core energy.
\end{abstract}

\begin{multicols}{2}
\narrowtext

\newcommand{\bc}{\begin{center}}
\newcommand{\ec}{\end{center}}
\newcommand{\be}{\begin{equation}}
\newcommand{\ee}{\end{equation}}
\newcommand{\beqn}{\begin{eqnarray}}
\newcommand{\eeqn}{\end{eqnarray}}
\newcommand{\ba}{\begin{array}}
\newcommand{\ea}{\end{array}}

\section{Introduction}
At low temperatures the physics of crystal surfaces on disordered substrates
is dominated by the randomness rather than thermal fluctuations.  In
2+1-dimensions this elastic surface is expected to have a roughening phase
transition at a critical temperature $T_c$ from a thermally {\it rough} phase
for $T > T_c$ to a {\it superrough} phase for $T<T_c $
\cite{TDV90,Sch95,TS94}, corresponding to a height-height correlation function
$\log(r)$ and $\log^2(r)$ respectively.  The $\log^2(r)$-superrough behavior
was numerically confirmed at finite temperature via Monte Carlo simulations
\cite{LR95} as well as in the limit of a vanishing temperature via exact
ground state calculation using combinatorial optimization
methods\cite{ZMS96,BHMR96,RB97}.

In this paper, we study the stability of the low-temperature (glassy) phase of
the solid-on-solid model (SOS) on a disordered
substrate\cite{RB97,Rie98a,Rie99} with respect to the formation of topological
point-like defects.  We also consider the density of defects and the screening
effect of pre-existing pairs to an introduced extra pair and allow for a
vortex-core energy.

\begin{figure}
\psfig{file=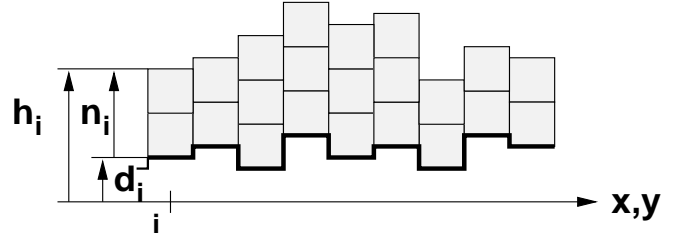,width=\columnwidth}
\caption{\label{height} \small
Height profile $h_i = n_i + d_i$ in the (2+1) SOS model, where $d_i \in [0,1]$ 
are the random offsets of the disordered substrate and $n_i$ the crystal 
heights (arbitrary integers) on the lattice site $i$. 
}
\end{figure}

\begin{figure}
\psfig{file=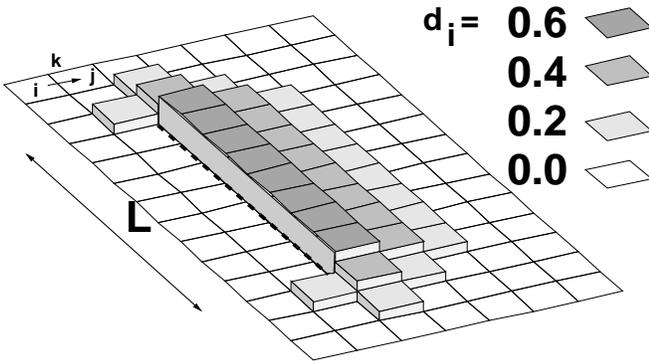,width=\columnwidth}
\caption{\label{disl_config} \small
Example of a disordered substrate heights $d_i$ in a random-surface model with
a single dislocation pair connected along a straight line of size $L$ (dashed
line). 
The optimal surface without dislocations would be flat, i.e. $n_i =0$ for all 
sites $i$, however, allowing dislocations would decrease the ground state 
energy (see text).
}
\end{figure}

The SOS model on a disordered substrate is given by the uniformly distributed
{\it substrate} height $d_i \in [0,1]$ and the integer {\it crystal} height
$n_i$ on a simple cubic $L\times L$-lattice $G$ with periodic b.c. and lattice
site $i$ as schematically shown in Fig.\ref{height}.  The $h_i = n_i + d_i$
denotes the {\it total surface} height at site $i$ and the SOS model
Hamiltonian is defined by
\be
H = \sum\limits_{\langle ij\rangle} (h_i - h_j)^2,
\label{SOS}
\ee
where the sum runs over all nearest-neighbor pairs $\langle ij\rangle$.  To
calculate the ground state of the SOS Hamiltonian (\ref{SOS}) we introduce the
{\it crystal} height-differences ${\bf n}^*_{ij} = n_i - n_j$ (integer) and
{\it substrate} height-differences ${\bf d}^*_{ij}=d_j-d_i$ ($\in[-1,+1]$)
along the links $k = (i,j)$ on the dual lattice
\noindent
$G^*$.
Thus we get the following SOS Hamiltonian for the dual space
\be
 H(\{ {\bf n}^*_k \}) = \sum\limits_k ({\bf n}^*_k - {\bf d}^*_k)^2 .
\label{SOS_k} \ee
\noindent
The minimal (optimal) energy configuration $\{ {\bf n}^*_k \}_{min}$ will just
be the closest integer ${\bf n}^*_k$ to ${\bf d}^*_k$ for all links $k=(i,j)$.
On the other hand, since the ${\bf n}^*_k$ describe height-differences in the 
scalar field given by the $n_i$ their sum along any oriented cycle on the 
surface around site $i$ has to be zero, i.e. the lattice divergence of 
${\bf n}^*$ has to vanish for each site $i$:
\be (\nabla \cdot {\bf n}^*)_i=0. \label{con}\ee
Note that $n_i$ can be considered to be a potential and ${\bf n}^*_{ij}$ as its
force field.
Obviously, for a typical disordered substrate the minimal configuration 
$\{ {\bf n}^*_k \}_{min}$ violates the mass balance constraint (\ref{con}).
Fig.\ref{disl_config} shows an example of a disordered substrate with 
substrate height $d_i =0.0$, 0.2, 0.4 and 0.6.
Consider the differences ${\bf d}^*_k$: across the dashed line we have 
${\bf d}^*_k =0.6$ and $|{\bf d}^*_k| < 0.5$ elsewhere.
Consequently, the absolute minimum-energy configuration without any balance 
constraint is given by  ${\bf n}^*_k = 1$ 
\noindent
and ${\bf n}^*_k = 0$ respectively.
With respect to the balance constraint (\ref{con}) the only feasible optimal 
solution (ground state) is a flat surface, i.e. ${\bf n}^*_k =0$ for all links
$k=(i,j)$.
On the other hand, dislocations of Burgers charge\cite{bc} $b$  can be 
introduced if one treats the height field $h_i$ as a multi-valued function 
which may jump by $b$ along lines that connect two point defects (i.e. a 
dislocation pair)\cite{CL97}.
Therefore, for the given example (Fig.\ref{disl_config}) it should be clear 
that the minimal configuration $\{ {\bf n}^*_k \}_{min}$ (see above) is exactly
the optimal (i.e. ground state) configuration with one dislocation pair.
One of the two defects has a Burgers charge $b=+1$ and the other one $b=-1$.
The pair is connected by a dislocation line (dashed line in Fig.
\ref{disl_config}) along which one has ${\bf n}^*_i = 1$.
This already demonstrates that due to the disorder the presence of dislocations
decreases the ground state energy and a proliferation of defects appears.
Alternatively, in Ref.\cite{RB97} a dislocation pair (excited step) was 
introduced by fixing the boundary to zero and one.

\section{Defect pairs in the SOS model}
The defect pairs in the disordered SOS model are source and sink nodes of
strength $+b$ and $-b$, respectively, for the network flow field $n_i$
\cite{Rie98a,Rie99}, which otherwise fulfills $(\nabla \cdot {\bf n}^*)_i =0$,
i.e. we have to modify the mass balance constraint (\ref{con}) as follows
\be
({\bf \nabla} \cdot {\bf n}^*)_i = \left\{ 
\begin{array}{r@{\quad , \quad}l}
0 & \mbox{no dislocation at} \; i\\
\pm b & \mbox{dislocation at} \; i
\end{array}
\right.
\label{SOS_con}
\ee
Thus the ground state problem is to minimize the Hamiltonian (\ref{SOS_k})
subjected to the mass balance constraint (\ref{SOS_con}) which can be solved by
the successive-shortest-path algorithm \cite{BHMR96,Rie98a,Rie99}.
In the following we concentrate on defect pairs with $b=\pm1$.

The defect energy $\Delta E$ is the difference of the minimal energy 
configuration {\it with} and {\it without} dislocations for each disorder 
realization, i.e. $\Delta E = E_1 - E_0$.
More
\begin{figure}
\psfig{file=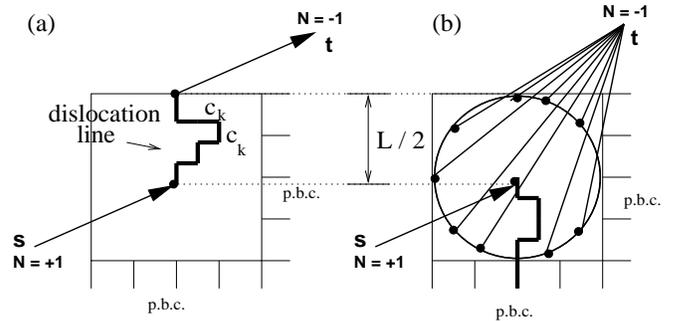,width=\columnwidth}
\caption{\label{topo}\small
Graph of a $L\times L$-lattice with periodic b.c. for the implementation
(a) of one {\it fixed} defect pair and (b) of a {\it partially optimized} pair.
Both are separated by $L/2$.
The energetic costs are $c_k({\bf n}^*_k) = ({\bf n}^*_k-{\bf d}^*_k)^2$ at the
dual site $k = (i,j)$. 
Dislocations are induced by two extra nodes $s$ and $t$, which are 
connected with the possible positions of the defects (big dots).
}
\end{figure}
\noindent
precisely, for the configuration {\it with} $N$ defect pairs of Burgers
charge $b=\pm1$ we introduce two extra nodes $s$ and $t$ with $n_s=+N$ and 
$n_t = -N$ respectively and connect them via external edges or bonds with 
particular sites of the lattice depending on the degree of optimization:
(a) with two sites separated by $L/2$ (Fig.\ref{topo}(a)), (b) the source node 
with one site $i$ and the sink node with the sites on a circle of radius $L/2$
around $i$ (Fig.\ref{topo}(b)) and (c) both nodes with the whole lattice.
Case (a) corresponds to a {\it fixed} defect pair, (b) to a 
{\it partially optimized} pair along a circle, both separated by a distance 
$L/2$, and (c) to a {\it completely optimized} pair with an arbitrary 
separation.
In all cases the energy costs for flow along these external edges are set to a
positive value in order to ensure the algorithm to find the optimal defect pair
on the chosen sites.
These ''costs'' have no contribution to the ground state energy. 
In case of {\it multi pairs} we always use graph (c).
Here, the optimal number $N$ of defects in the system is gradually determined 
starting with one pair ($N=1$) with a vortex core energy $2E_c$ and checking 
whether there is an energy gain or not.
If yes, add a further pair (with $2E_c$) and repeat the procedure until there 
is no energy gain from the difference of the ground state energy between two 
iterations.

\section{Single defect pair ($N=1$) \label{N=1}} 
\begin{figure}
\psfig{file=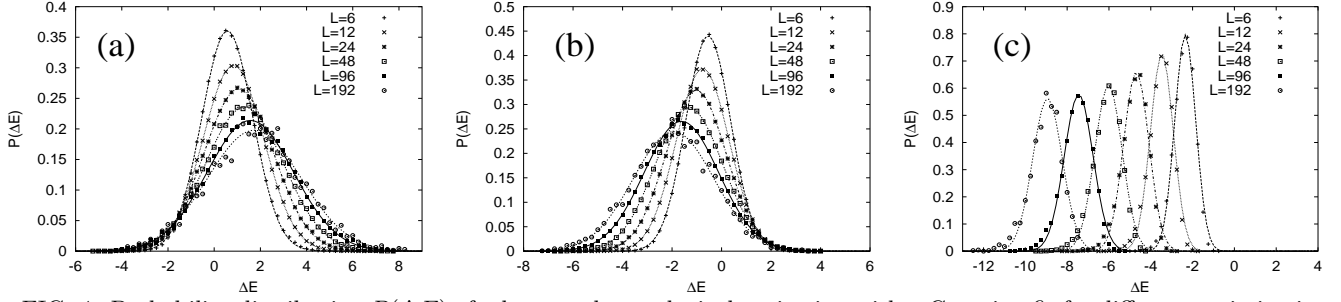,width=\columnwidth}
\caption{\label{dE}\small
Probability distribution $P(\Delta E)$ of a large-scale topological excitation
with a Gaussian fit for different optimizations: (a) for a {\it fixed} defect 
pair, (b) for a {\it partially optimized} pair and (c) for a 
{\it completely optimized} pair with different system sizes $L$.
}
\end{figure}

We study the defect energy $\Delta E$ and its probability distribution $P(\Delta E)$ on a $L \times L$ lattice with
$L = 6$, 12, 24, 48, 96 and 192 and $2\cdot 10^3 - 10^5$ samples for
each size and consider the three cases (a)-(c) (see above).
With an increasing degree of optimization a negative defect energy $\Delta E$
becomes more probable and its probability distribution $P(\Delta E)$ differs 
more and more from the Gaussian fit, Fig.\ref{dE}.
The resulting disorder averaged defect energy $[\Delta E]_{dis}$ scales like

\end{multicols}
\widetext

\begin{multicols}{2}
\newcommand{\bc}{\begin{center}}
\newcommand{\ec}{\end{center}}
\newcommand{\be}{\begin{equation}}
\newcommand{\ee}{\end{equation}}
\newcommand{\beqn}{\begin{eqnarray}}
\newcommand{\eeqn}{\end{eqnarray}}
\newcommand{\ba}{\begin{array}}
\newcommand{\ea}{\end{array}}
\noindent

\be
[\Delta E]_{dis} \sim \left\{ 
\begin{array}{c@{, \;}l}
\ln (L) & \mbox{\small fixed defect pair}\\
-0.27(7)\cdot\ln^{3/2}(L) & \mbox{\small partially optimized}\\
-0.73(8)\cdot\ln^{3/2}(L) & \mbox{\small completely optimized}
\end{array}
\right.
\label{dE_eq}
\ee
and its related variance $\sigma$ like
\be
\sigma(\Delta E) \sim \left\{ 
\begin{array}{c@{, \quad}l}
\ln (L) & \mbox{\small fixed defect pair}\\
\ln^{2/3}(L) & \mbox{\small partially optimized}\\
\ln^{1/2}(L) & \mbox{\small completely optimized}
\end{array}
\right.
\ee
where the exponents are approximations for the best data collapse.
The defect energy indicates that for the optimized cases dislocations can
proliferate due to thermal fluctuations and melt the elastic superrough phase.
Furthermore, for a growing degree of optimization the scaling amplitude of
$[\Delta E]_{dis}$ increases.

The mean length (mass) $l_{DL}$ of the line connecting the two defects scales 
with the system size $L$ according to the fractal dimension 
\be
d_f = 1.28 \pm 0.02
\label{df_simgle}
\ee
for the {\it fixed} and {\it partially optimized} situation.
\begin{figure}
\psfig{file=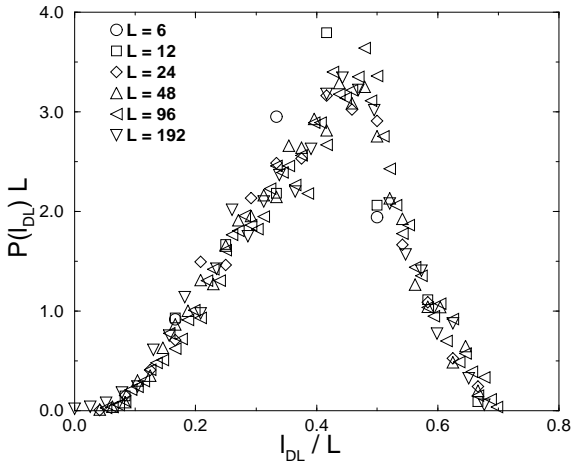,width=\columnwidth}
\caption{\label{P_l}\small
Finite-size-scaling relation of the probability distribution
$P(l_{DL})$ of the mean distance $l_{DL}$ between two optimally placed 
dislocations for system size $L = 6$, 12, 24, 48, 96 and 192. 
The data collapse for $P_L(l_{DL}) \sim 1/L \cdot p(l_{DL} / L)$.}
\end{figure}
\noindent
For the {\it completely optimized} case Fig.\ref{P_l} shows a probability 
distribution $P(l_{DL})$, which behaves like
\be
P_L(l_{DL}) \sim \frac{1}{L} \cdot p\left(\frac{l_{DL}}{L}\right).
\ee

\section{Multi defect pairs ($N>1$)}
\begin{figure}
\psfig{file=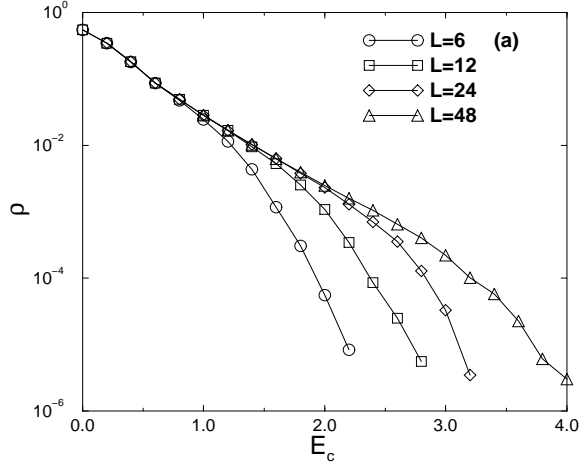,width=\columnwidth}
\psfig{file=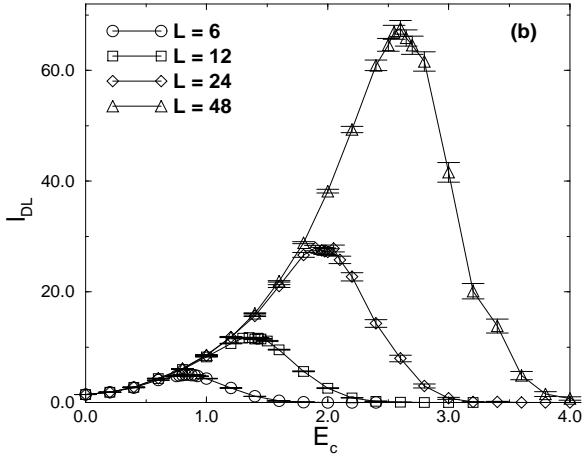,width=\columnwidth}
\caption{\label{E_c}\small
(a) Density $\rho$ of defects with respect to the vortex core energy $E_c$ for
different system sizes $L$ = 6 - 48 and $10^3$ up to $10^4$ samples.
The log-lin plot indicates an exponential decay of $\rho$.
Simultaneously, (b) the mean distance $l_{DL}$ of all dislocation pairs 
vs. the vortex core energy $E_c$.
Comparing (a) and (b) one sees that the maximal lengths $l_{DL}$ occurs 
at the cross-over energy $E_c^{max}$ (see text).
}
\end{figure}

Next, we study the effect of a uniformly given vortex-core energy $E_c$ to 
the system of {\it multi defect pairs} ($N > 1$) as a simplification of the 
real situation with a distribution of $E_c$.
As shown in Fig.\ref{E_c}(a), the density $\rho$ of defects decays 
exponentially with an increasing $E_c$, i.e.
\be
\label{rho}
\rho(E_c) \sim \mbox{e}^{(- E_c / E_0)^\alpha}.
\ee
For the $E_0$ and $\alpha$ we can distinguish between two intervals of $E_c$
which refers to a stretched and a pure exponential decay, respectively.
In detail we have

\bc
\begin{tabular}{c@{\qquad}c@{\qquad}c} \hline
$E_c \in$ & $E_0$ & $\alpha$\\ \hline
$[0,\infty[$ & $0.75 \pm 0.15$ & $0.6 \pm0.2$ \\[0.2cm]
$[0,E_c^{max}(L)[$ & $ 0.45 \pm 0.03$ & 1 \\[0.2cm] \hline
\end{tabular}
\ec
The upper limit ${E_c}^{max}(L)$ corresponds to the maximal mean length 
$l_{DL}$ for each system size $L$, c.f. Fig.\ref{E_c}(a) and (b), and scales 
like
\be
\label{E_max}
{E_c}^{max} \approx (\mbox{const.} +0.47 \pm 0.02)\cdot\ln(L))^{3/2}.
\ee
Moreover, we found the same scaling behavior for the vanishing defect energy, 
i.e. $[\Delta E]_{dis} = 0$:
\be
\label{E_dE}
E_{c0} \approx (\mbox{const.} + (0.47 \pm 0.01)\cdot\ln(L))^{3/2}.
\ee
From the plot of the maximal mean length $l_{DL}$ (Fig.\ref{E_c}(b)) vs. the 
system size $L$, i.e. $l_{DL}(E_c^{max}) \sim L^{d_f}$, the fractal dimension 
$d_f$ is given by 
\be
d_f = 1.267 \pm 0.07,
\ee
\noindent
close to the one of the above single line situation 
($d_f \approx 1.28$).

\begin{figure}
\psfig{file=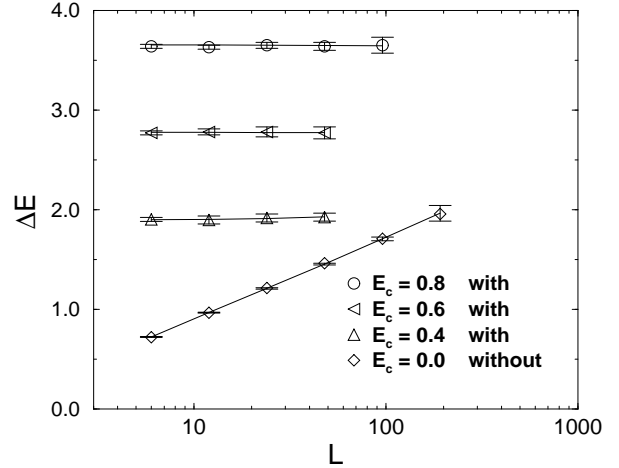,width=\columnwidth}
\caption{\label{screen}\small
Defect energy $\Delta E$ of a single (introduced) defect pair vs. the system 
size $L$ ($L=6$, 12, 24, 48, 96) in a system {\it with} and {\it without} an 
already optimal number of dislocations for different vortex core energies 
$E_c$.
}
\end{figure}

Finally, we focus on the effect of introducing an {\it extra} fixed defect 
pair separated by $L/2$ to an already (completely) optimized configuration 
with a vortex core energy $E_c$.
This extra pair costs
\be
\Delta E_{fix} = E'_1 + 2 E_c - E_1,
\ee
where $E_1$ denotes the ground state energy for $N$ (pre-existing optimal) 
pairs and $E'_1$ the energy for $N+1$ optimally placed pairs, both for the 
same disorder configuration $\{d_i \}$.
As plotted in Fig.\ref{screen}, $\Delta E_{fix}$ is constant in $L$, but
linear in $E_c$, i.e.
\be
\Delta E_{fix}(L) = (0.17 \pm 0.02) + (4.35 \pm 0.02 )\cdot E_c.
\label{E_fix}
\ee
Thus, one obtains a screening effect of the defect-defect interaction due to 
disorder-induced dislocations.
In comparison, Fig.\ref{screen} also shows the case for a single pair
($N=1$) {\it without} pre-existing pairs as studied in section \ref{N=1}.

\section{Related models}
A similar picture of the effect of dislocations to a randomly pinned elastic
media at $T=0$ were found for other discrete models, the {\it fully-packed
  loop} (FPL) model\cite{ZLF99} and the matching model\cite{Mid98}, both on a
bipartite hexagonal lattices with a linear energetic cost function and
periodic b.c.

In the case of a single {\it fixed} defect pair we found the same
$\ln(L)$-behavior of the defect energy as for the excitation step in Ref.
\cite{RB97,ZLF99}, but got a smaller fractal dimension $d_f= 1.28(2)$ rather
than $d_f=1.35(2)$\cite{RB97}.  The disorder-induced dislocations turned out
to destroy the quasi-long-range order of the elastic phase due to a negative
scaling behavior of defect energy $\Delta E$ with respect to $L$ for optimally
placed defects, i.e.  $\Delta E \sim -\ln^{3/2}(L)$, in good agreement with
the results of the FPL model\cite{ZLF99}.  When taking into account screening
and an uniform vortex core energy $E_c$ in addition to the energy balance
$\Delta E$, one finds that the energetic costs $\Delta E_{fix}$ of an
introduced fixed pair does not depend on the system size $L$, as shown for the
FPL\cite{ZL98} and matching \cite{Mid98} model.  In addition we found for the
disordered SOS model a linear dependence of $\Delta E_{fix}$ on the vortex
core energy $E_c$ (Fig. \ref{screen}).  One concludes that this extra pair is
screened by the pre-existing defect pairs.  For the exponentially decay of
density $\rho$ of the dislocation pairs we distinguished between (a) the whole
range of the vortex core energy $E_c$ and (b) a range with an upper limit
$E_c^{max}(L)$, for which (latter case) the mean defect length $l_{DL}$ was
found to be maximal.  The case (a) corresponds to $\alpha = 1$ and $E_0 =
0.45(3)$ and the maximal length $l_{DL}$ related to the cross-over energy
$E_C^{max}(L)$ behaves as $l_{DL} \sim L^{d_f}$ with the fractal dimension
$d_f = 1.267(7)$.  Both results were also found in Ref.\cite{Mid98}.  For the
case (b) we get $\alpha \approx 0.75$ (close to $2/3$) and $E_0 \approx 0.6$
in good agreement with Ref.\cite{ZL98}.

Finally, we relate the SOS model (\ref{SOS}) to the continuum description of a
randomly pinned elastic medium on large length scales given by the {\it
  sine-Gordon} model Hamiltonian
\be
H= \int \mbox{d}^2{\bf r}\left[\frac{K}{2}(\nabla u({\bf r}))^2 
- w \cos\left(2\pi (u({\bf r}) - d({\bf r})) \right)\right]
\label{sgm}
\ee
where $K$ is the elastic constant and $d({\bf r})$ a random field out of 
[0,1].
The first term represents the elastic energy $E_{el}$ and the second one the 
random pinning energy $E_{pin}$.
The model is known to describe a weakly disturbed vortex lattice in a thin 
two-dimensional (2D) superconducting film introduced by a parallel field
\cite{Bla94,GD98,NS99}.
Other experimental realizations are charge density waves\cite{Gru88} and Wigner
crystallization of electrons\cite{Li98}.

The relation to the SOS model is as follows: in the limit of an infinite
coupling strength $w \to \infty$ and $T=0$ the {\it sine-Gordon} model maps
onto a lattice SOS model Eq.(\ref{SOS}), as the cosine-term of Eq.(\ref{sgm})
forces the displacement field $u({\bf r})$ to be $u({\bf r}) = d({\bf r}) +
n({\bf r})$\cite{LR95,Rie98,Rie99}, where $n({\bf r})$ is an integer.  One can
identify the $u({\bf r})$ as the continuous height field $h({\bf r})$ of the
SOS model.

The results of the analytical study of the {\it sine-Gordon} model
\cite{Fis97,ZLF99,DG98} are in good agreement with our results, but only refer
to the cases of {\it fixed} and {\it completely optimized} pairs.
Furthermore, these studies allow another interpretation of the defect energy
$\Delta E$ and density $\rho$.  From the calculation of the elastic energy
$E_{el}$ \cite{Kos74} and defect energy $\Delta E$ \cite{DG98,ZLF99,Fis97} one
gets that for a {\it fixed} pair the elastic energy $E_{el}$ dominates the
pinning energy $E_{pin}$, i.e.  $\Delta E \sim E_{el}$, and for the {\it
  completely optimized} pair the situation is vice versa, i.e. $\Delta E \sim
E_{pin}$.  The resulting scaling behavior is found to be $\Delta E \sim
\ln(L)$ and $\Delta E \sim -\ln^{3/2}(L)$, respectively \cite{ZLF99}.  The
scaling behavior of the {\it fixed} dislocation pair in presence of pinning
disorder is essentially equivalent to the one of a {\it fixed} defect pair at
finite temperatures $T$ without disorder, i.e. $\Delta E \sim \ln(L) \sim
E_{el}^{pure}(T)$.

The density $\rho$ of defects can be related to the length scale $\xi_D$
beyond which the dislocations become unpaired\cite{DG98} since for $\alpha
\approx 0.6$ the density $\rho$, Eq. (\ref{rho}), shows the same scaling
behavior as $\xi_D$ in the case of low temperatures and large core energy
$E_c$, i.e. $E_c \gg K \ln(\xi_D)$.  For $E_c \approx 0$ we found large
densities $\rho$ and one is probably out of the regime given by $E_c \gg K
\ln(\xi_D)$.  This would possibly explain the occurrence of the stretched
exponential behavior close to $E_c \approx 0$ as seen in Fig.\ref{E_c}(a).

To summarize we have studied the effect of dislocation pairs on the ground
state properties of the SOS model on a disordered substrate. For a fixed
position of the dislocation pair a distance $L$ apart we found that on average
t he defect {\it costs} an energy proportional to $\ln L$, in agreement with
the findings for the energy costs for a steplike excitation step with fixed
end points reported earlier \cite{RB97} and also in agreement with recent
results for other two-dimensional lattice models \cite{ZLF99,Mid98}. On the
other hand if we optimize the position of the dislocation pair we showed that
it {\it gains} energy, namely an amount proportional to $\ln^\psi L$ with an
exponent $\psi$ around $3/2$ as predicted by scaling arguments and also
observed in the FPL model \cite{ZLF99}. When introducing a penalty for the
topological defects (i.e.\ a core energy) we showed that the density of
defects vanishes exponentially as a function of this core energy, which is in
agreement with the results for the FPL model \cite{Mid98}. Finally we also
demonstrated that a dislocation pair is screened by the presence of other
dislocations in the system.

\end{multicols}

\end{document}